\documentclass[aps,prl,showpacs,notitlepage,twocolumn,superscriptaddress,nofootinbib,preprintnumbers]{revtex4-1}

\usepackage{graphicx}      
\usepackage{bm}            
\usepackage[colorlinks=true]{hyperref}  
\usepackage{amssymb}
\usepackage{amsmath}
\usepackage{physics}
\usepackage{tabularx}

\newcommand{\half}{\dfrac{1}{2}}
\newcommand{\up}{\uparrow}
\newcommand{\down}{\downarrow}

\begin{document}
\title{Quantum Computation and Simulation using Fermion-Pair Registers}
\author{Xiangkai Sun}
\affiliation{Department of Physics, Massachusetts Institute of Technology, Cambridge, MA 02139, USA}
\author{Di Luo}
\affiliation{Center for Theoretical Physics, Massachusetts Institute of Technology, Cambridge, MA 02139, USA}
\affiliation{The NSF AI Institute for Artificial Intelligence and Fundamental Interactions}
\affiliation{Department of Physics, Harvard University, Cambridge, MA 02138, USA}
\author{Soonwon Choi}
\affiliation{Center for Theoretical Physics, Massachusetts Institute of Technology, Cambridge, MA 02139, USA}

\date{\today}

\preprint{MIT-CTP/5559}

\begin{abstract}

We propose and analyze an approach to realize quantum computation and simulation using fermionic particles under quantum gas microscopes.
Our work is inspired by a recent experimental demonstration of large-scale quantum registers, where tightly localized fermion pairs are used to encode qubits exhibiting long coherence time and robustness against laser intensity noise.
We describe how to engineer the SWAP gate and high-fidelity controlled-phase gates by adjusting the fermion hopping as well as Feshbach interaction strengths.
Combined with previously demonstrated single-qubit rotations, these gates establish the computational universality of the system.
Furthermore, we show that 2D quantum Ising Hamiltonians with tunable transverse and longitudinal fields can be efficient simulated by modulating Feshbach interaction strengths.
We present a sample-efficient protocol to characterize engineered gates and Hamiltonian dynamics based on an improved classical shadow process tomography that requires minimal experimental controls.
Our work opens up new opportunities to harness existing ultracold quantum gases for quantum information sciences.
\end{abstract}

\maketitle

\textit{Introduction.} 
Quantum gas microscopes~\cite{bakr2009quantum,sherson2010single,cheuk2015quantum} provide a powerful toolbox for probing quantum many-body physics. A number of interesting many-body quantum phenomena have been investigated through quantum simulations on the platform, such as Pauli blocking~\cite{omran2015microscopic}, Mott insulators~\cite{greif2016site,cheuk2015quantum}, many-body localization~\cite{choi2016exploring,rubio2019many}, bad metallic transport~\cite{brown2019bad}, and subdiffusion and transport in Hubbard model~\cite{guardado2020subdiffusion,Nichols_2019}. Besides the physics applications, a few early attempts for quantum information processing have also been made, including quantum walk~\cite{preiss2015strongly} and entanglement detection~\cite{islam2015measuring}.

Recently, exciting progress has been reported on realizing a large-scale quantum register of fermion pairs with a quantum gas microscope~\cite{Hartke_2022}, where qubits are engineered from the common and relative motional degrees of freedom of fermion pairs and protected by the fermionic exchange symmetry. The fermion pair quantum register is shown to exhibit a long-lived coherence time and robustness to the experimental noise in the confining potential. It is scalable, programmable with different geometries~\cite{greif2013short,yang2021site}, and flexible with individual controls~\cite{sherson2010single,haller2015single,parsons2015site}. While this development provides promising opportunities for quantum information sciences, an important question remains open: how to perform universal quantum computation and many-body simulations through the platform. 

In this paper, we address the above question and develop schemes for programmable universal quantum computation and quantum many-body simulation using a quantum register of fermion pairs. We propose how to engineer two-qubit gates, including the SWAP gate and controlled-phase gates. These are achieved by constructing effective Hamiltonians by tuning the nearest-neighbor hopping $J$ and the Feshbach interaction $U$ in the fermionic system. 
Since the global single qubit rotation has been demonstrated~\cite{Hartke_2022} and individual addressing is feasible~\cite{edge2015imaging,weitenberg2011single,zupancic2016ultra,wang2016single}, the realization of entangling two-qubit gates completes a universal gate set. Furthermore, we develop a scheme based on modulating $U$ for realizing 2D quantum Ising model simulations with tunable transverse and longitudinal fields in a programmable geometry. To characterize the quantum gates and the quantum dynamics in experiments, we present a classical shadow tomography protocol for quantum processes, optimized for the existing experimental capability and sample-efficient with theoretical guarantee. Our approach opens up new possibilities for quantum computation and simulation with quantum gas microscope technologies.

\begin{figure}[t!]
		\centering
		\includegraphics[width=\columnwidth]{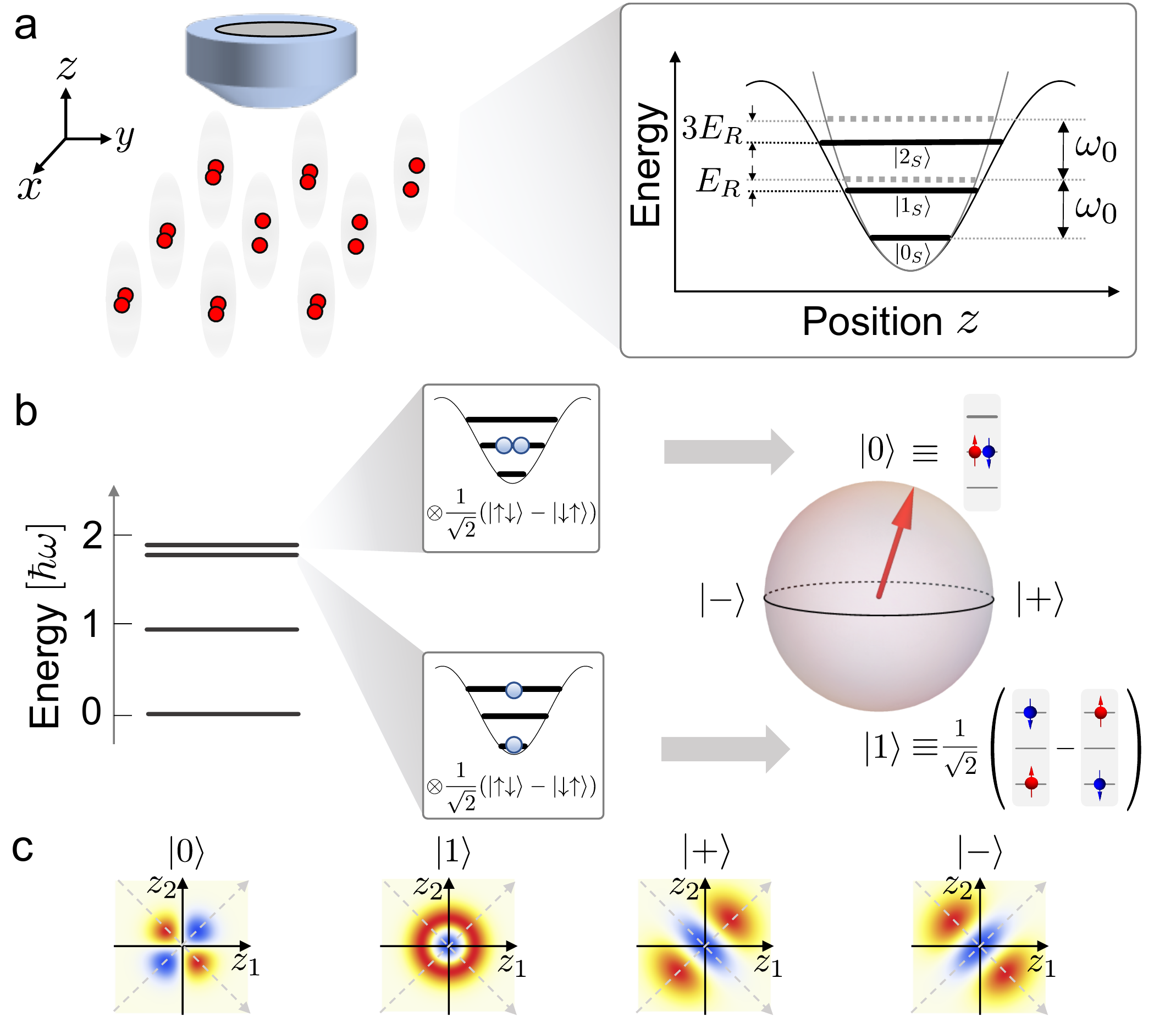}
		\caption{(a) The fermion pair register in a quantum gas microscope setup. The register is an optical lattice with two fermions per site, of which vibrational states in the $z$ direction encode a qubit. Each site can be viewed as an 1-dimensional anharmonic trap. The energy levels of the single particles show that the anharmonicity of the $z$ potential results in uneven splitting between the first three vibrational states. (b) The qubit encoding scheme represented on a Bloch sphere. Each of the qubit states $\ket{0}$ and $\ket{1}$ is encoded by a tensor product of a spin singlet and a spatially symmetric vibrational state. The anharmonicity leads to a stable energy difference $E_R$ between the two qubit states. (c) The two-body wavefunctions $\psi(z_1, z_2)$ of 4 single-qubit states.}
        \label{f1}
	\end{figure}

\textit{Fermion-pair registers.} We start by briefly reviewing how to encode qubits in the fermion-pair register~\cite{Hartke_2022}. Figure \ref{f1}(a) shows the schematic of pairs of fermionic atoms trapped in every site of an optical lattice. Qubits are encoded by the vibrational states of a fermion pair in the slightly anharmonic symmetric trap in one dimension.
The logical states $\ket{0}$ and $\ket{1}$ are formed by spin singlet states, in which $\ket{0}$ has both particles in the first excited state, whereas $\ket{1}$ has one particle in the second excited state.

Both qubit states carry approximately two energy quanta $2\hbar \omega_0$, but their energies differ by $E_R$, the recoil energy, due to the anharmonicity.
These states span the single-qubit subspace $\mathcal{H}_L$.

As already demonstrated~\cite{Hartke_2022}, qubits can be robustly manipulated by modulating a magnetic field at angular frequency $E_R$.
This is because the interaction between the pair $\hat{U}$ can be tuned via Feshbach resonance and depends on the distance between two atoms, $z_r = z_1-z_2$, which distinguishes two quantum states in $\mathcal{H}_L$. Specifically, the Feshbach interaction $\hat{U} = U\, a_z\, \delta(\hat{z}_1-\hat{z}_2)$, where $a_z$ is the optical lattice periodicity in the $z$ direction, and $U$ is the interaction strength tunable by an external magnetic field. This interaction is shown to be diagonal in the $\ket{\pm} = (\ket{0} \pm \ket{1})/\sqrt{2}$ basis, serving as a $\sigma_x$ term in the computational basis (see Figure~\ref{f1}(c))~\cite{Hartke_2022}.

\textit{Engineering gates.}
To show that one can engineer a computationally universal gate set, we derive an effective unitary dynamics and Hamiltonian for qubits, starting with the microscopic description of fermions in an optical lattice.
In particular, we demonstrate the implementation of the SWAP gate $\ket{\psi_1}\otimes \ket{\psi_2} \mapsto \ket{\psi_2}\otimes \ket{\psi_1}$ as well as the CPHASE gate $\mathrm{CP}_\phi = \textrm{diag}(1,1,1,e^{i\phi})$ in the computational basis.

We consider the dynamics of fermions described by the Fermi-Hubbard model $H = H_E + H_U + H_J$, where each term describes the orbital energy, the on-site interaction, and the nearest-neighbor tunneling, respectively. Specifically, the orbital energy term $H_E$ describes the motion of fermions in a slightly anharmonic trap. The energy spectrum of a single fermion resembles that of a simple harmonic oscillator of angular frequency $\omega_0$ dependent on the lattice depth, while the anharmonicity introduces nonlinear shifts to the eigenenergies, insensitive to the lattice depth. The energy difference between the ground state and the first excited state is reduced by the recoil energy $E_R = \pi^2 \hbar^2 / (2 m a_z^2)$ (Figure~\ref{f1}(a)), which only depends on the atomic mass $m$ and the lattice periodicity $a_z$. 
The interaction $H_U$ describes the two-body $s$-wave Feshbach interaction between atoms, proportional to the interaction strength $U$, which couples different orbitals. In the deep potential limit ($\hbar\omega_0 \gg E_R \gg U$), the interaction is effectively on-site because that between particles in different sites is strongly suppressed due to exponentially decaying spatial wavefunctions. 
The tunneling term $H_J$ describes single particle hopping between adjacent sites in the $xy$-plane. The hopping amplitude $J$ is independent of the energy levels associated with the harmonic motion along $z$-direction because all particles are assumed to be in the motional ground of $xy$-plane (see Sec.I of \cite{sup}).

The Hamiltonian preserves total spin, and in the limit that the energy splittings $\omega_0$ are much greater than other energy scales, $\hbar \omega_0 \gg E_R,U, J$, all dynamics that do not preserve the number of energy quanta $\omega_0$ are negligible. In the following discussion, we isolate two adjacent sites from the environment to calculate the nearest-neighbor effective Hamiltonian and engineer two-qubit gates. The interaction $U$ and the hopping $J$ are both tunable. 

\begin{figure}[t!]
	\centering
	\includegraphics[width=\columnwidth]{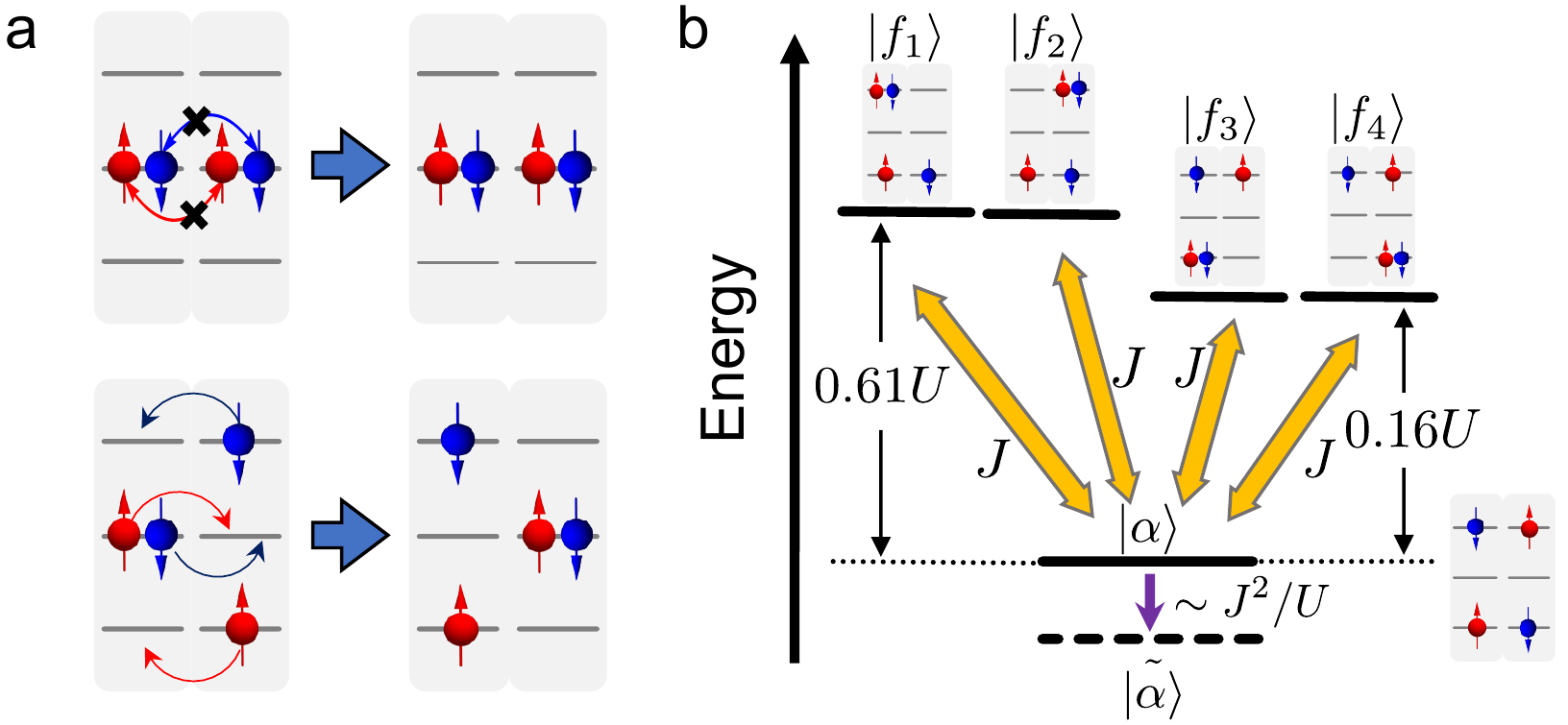}
	\caption{(a) Two hopping processes that are involved in the SWAP gate implementation. Top: a hopping process that is forbidden due to the Pauli exclusion principle, which leaves $\ket{00}$ unchanged. Bottom: one of the hopping processes that is allowed and involved in the exchange of $\ket{01}$ and $\ket{10}$. (b) Second-order tunneling processes that shift the energy of $\ket{\alpha}$, one component of the state $\ket{11}$. The energy of $\ket{\alpha}$ is lowered when the interaction between atoms is attractive, and the eigenstate is perturbed to $\tilde{\ket{\alpha}}$.}
        \label{f2}
\end{figure}

First, we note that the SWAP gate can be easily engineered.
In the absence of $U$, each atom can tunnel to its neighboring site freely with hopping amplitude $J$, which is pictorially illustrated in Figure~\ref{f2}(a). Atoms hop between the same energy levels with the same tunnelling amplitude, and their spins are preserved. For the initial state being $\ket{00}$, in which case the first excited states in both wells are fully occupied, the tunneling is forbidden by the Pauli exclusion principle. Starting from a product state $\ket{\psi_A}_L\otimes \ket{\psi_B}_R$ and evolving this system for $t = (2m+1)\pi / (2J)\ (m\in \mathbb{N})$, the system ends up in the state $\ket{\psi_B}_L\otimes \ket{\psi_A}_R$, which is the dynamics of a 2-qubit SWAP gate.
The SWAP gate is particularly useful, as it allows one to apply two-qubit gates on any arbitrary qubit pairs by first applying a sequence of SWAP gates to bring the pair next to each other before applying a desired gate.

\begin{figure}[t!]
	\centering
	\includegraphics[width=\columnwidth]{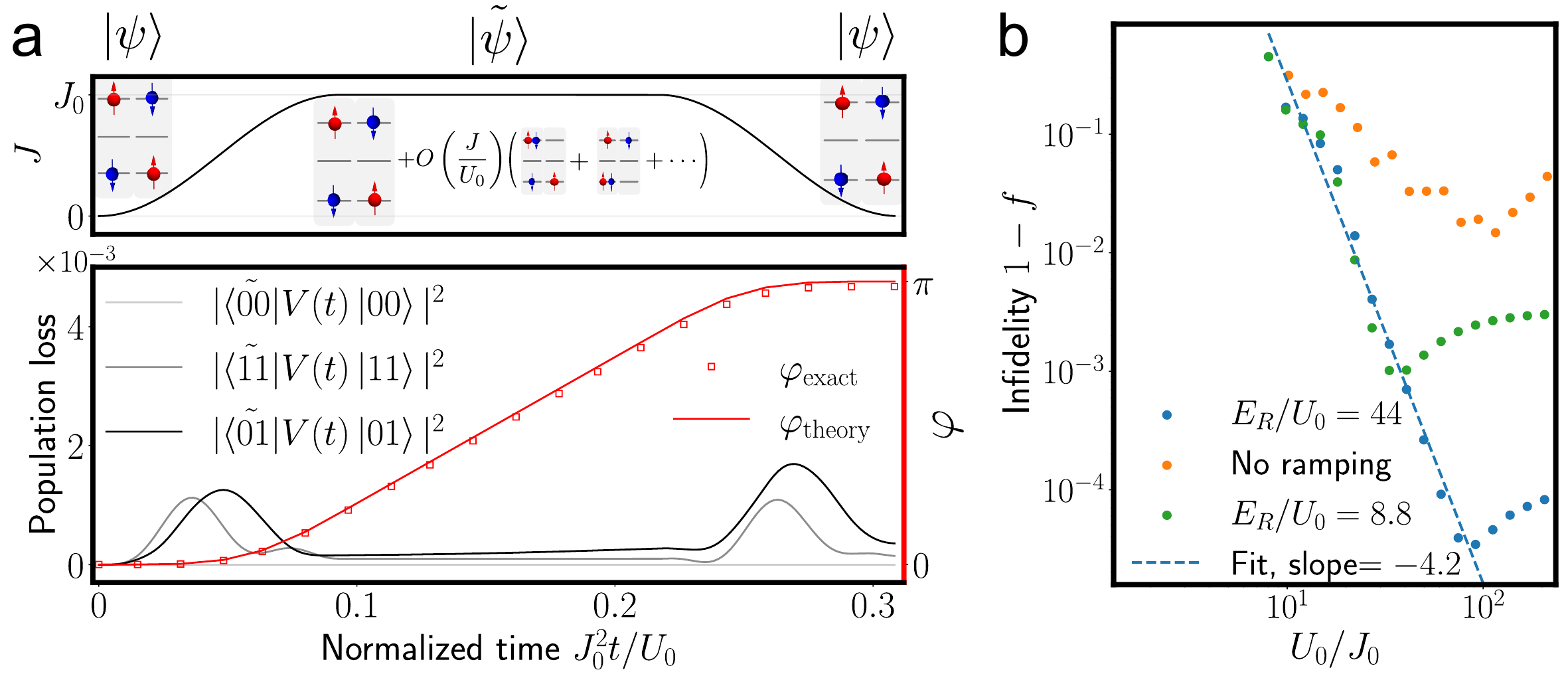}
	\caption{(a) Top: the adiabatic tuning protocol and dressed states.
 As the hopping strength $J$ is tuned from 0 to its maximum $J_0$ and back to 0 adiabatically,
a two-qubit state $\ket{\psi}$ in the Hilbert space $\mathcal{H}_L^{\otimes 2}$ evolves into a dressed state $\tilde{\ket{\psi}}$ under an effective interacting Hamiltonian in a dressed Hilbert space, and back to $\mathcal{H}_L^{\otimes 2}$.
The figure illustrates the evolution of a particular initial state as an example. 
Bottom: the controlled-Z gate (CZ) simulation with $U_0/J_0 = 30$. The parameters in the CZ protocol are chosen such that the induced phase $\varphi_{\mathrm{theory}}$ derived from the second order perturbation equals $\pi$ in the end. The phase shift $\varphi_{\mathrm{exact}}$ obtained from Fermi-Hubbard model simulation shows good agreement with $\varphi_{\mathrm{theory}}$. The population changes of $\ket{00},\ket{11},\ket{01}$ are on the order $10^{-3}$ during the CZ gate protocol, which are consistent with the goal of zero change of the adiabatic tuning. 
    (b) The infidelity between an ideal CZ gate and the exact time-evolution of  four fermions obtained from {\it ab initio} numerical simulations of the Fermi-Hubbard model~\cite{sup}, averaged over Haar random two-qubit initial states, as a function $U_0/J_0$ for two different values of $E_R/U_0$ with (blue and green) and without (yellow) adiabatic control of $J(t)$. With the adiabatic control, the infidelity scales as $O((J_0/U_0)^4)$ (blue and green).
    }
        \label{f3}
\end{figure}

Next, we show that CPHASE gates can be realized by perturbatively turning on $J$ and $U$ in the presence of large $E_R$ with the hierarchy of energy scales $E_R \gg U \gg J$, which will lead to an effective Ising interactions. The energy eigenstates of the leading-order Hamiltonian are $\{\ket{00}, \ket{01}, \ket{10}, \ket{11}\}$, among which two states $\ket{01}$ and $\ket{10}$ are degenerate. 
During gate implementation, the interaction strength $U(t)$, which is set to be constant $U_0$ at any time $t$, dresses local qubit states $\ket{0}$ and $\ket{1}$. The presence of $J$ introduces a coupling between two qubits via virtual processes (Figure~\ref{f2}(b)) and, up to the second order, perturbatively dresses the qubit states and shifts their energies, so that the energy difference between $\ket{00}$ and $\ket{01}$($\ket{10}$), denoted as $\omega-g$, is different from that between $\ket{01}$($\ket{10}$) and $\ket{11}$, denoted as $\omega+g\ (g\neq 0)$~\cite{sup}. We obtain the coupling coefficient $g \approx 16.2\ J^2/U_0$ by summing up all second-order virtual processes (see \cite{sup} for detailed derivation).
Therefore, the effect can be described by the following Hamiltonian
\begin{equation}\label{Cphaseham}
    H_{\mathrm{eff}} =  \frac{1}{2} 
 \omega(\sigma_z^{(1)}+\sigma_z^{(2)}) - \frac{1}{2}g \sigma_z^{(1)}\sigma_z^{(2)},
\end{equation}
in the dressed computational basis $\{\tilde{\ket{00}},\tilde{\ket{01}},\tilde{\ket{10}},\tilde{\ket{11}}\}$. 
Upon time evolution, we define the induced phase as $\varphi \equiv \varphi_{\ket{00}} + \varphi_{\ket{11}} - \varphi_{\ket{01}} - \varphi_{\ket{10}}$, where $\varphi_{\ket{\psi}} \equiv \mathrm{arg}{\tilde{\bra{\psi}} e^{-\mathrm{i}\int \dd t H(t)} \ket{\psi}}$ is the accumulated phase on state $\ket{\psi}$ in time $t$ from a time reference, and $\tilde{\ket{\psi}}$ is the eigenstate of $H(t)$ which depends on $J$ and $U$. By appropriately choosing the evolution time $T$ as well as parameters $J$ and $U$, one can implement the controlled-phase gate with the induced phase $\varphi$ up to single-qubit rotations~\cite{sup} (Figure~\ref{f3}(a)).

\textit{Fidelities.} In our approach, the fidelity of CPHASE is limited by two factors: (i) the difference between the encoded quantum states and their dressed versions and (ii) higher-order corrections in $J/U$, which may result in corrections in $\varphi$ and new terms in Eq.~\ref{Cphaseham}. The effect of (i) can be efficiently suppressed by adiabatically turning on and off $J(t)$ from $0$ to $J_0$ and back to $0$. That of (ii) on $\varphi$ can be eliminated by carefully calibrating $\varphi$ (e.g. adjusting evolution time) using the process tomography as we discuss further below. Here, we define the gate fidelity $f$ as the average fidelity between a Haar random initial state evolving under a CPHASE gate and the same state evolving under the Fermi-Hubbard model.
Ultimately, the fidelity is limited by the additionally introduced spin-exchange interactions $(\sigma_x^{(1)}\sigma_x^{(2)} + \sigma_y^{(1)}\sigma_y^{(2)})$ of order $O(J^4/U^3)$, resulting in an infidelity scaling as $O((J_0/U_0)^4)$~\cite{sup} (Figure~\ref{f3}(b)). 
The limited coherence time of qubits associated with light scattering prevents one from choosing arbitrarily small $J$. A reasonable choice of $U_0/J_0$ is around 22 for current technology~\cite{Hartke_2022}, which implies a fidelity of $99\%$.

\textit{2D Ising Model Simulation.}
The technique to engineer the effective Hamiltonian in Eq.~\ref{Cphaseham} for CPHASE gates can be further generalized to engineer the 2D transverse field Ising model
\begin{equation}\label{Isingham}
    H_{\mathrm{eff}} =  \sum_i \frac{1}{2} 
 ((\omega'-\Delta) \sigma_z^{i} + \Omega \sigma_x^{i}) - \sum_{i,j}\frac{1}{2} g_{i, j} \sigma_z^{i}\sigma_z^{j},
\end{equation}
Compared to Eq.~\eqref{Cphaseham}, the only difference is the addition of the transverse field $\Omega \sigma_x^i$, which can be realized by modulating the interaction strength $U(t) = U_0 + U_1 \cos ((E_R + \Delta)t)$. In the regime $U_0 \gg U_1$, the perturbative analysis discussed for the CPHASE gate still holds, leading to the diagonal terms in Eq.~\eqref{Isingham}.
The additional time-dependent matrix elements of $U$ modulated at $E_R + \Delta$  couple the two qubit states $\ket{0}$ and $\ket{1}$ off-resonantly with the detuning $\Delta \ll E_R$. Consequently, it introduces $\sigma_x^i$ with the Rabi frequency $\Omega = \sqrt{\pi/128}\ U_1$ and an energy shift $\omega' - \Delta$ on $\sigma_z^i$ in the interaction picture, where $\omega'$ is a geometry-dependent constant~\cite{sup}.

\begin{figure}[t!]
	\centering
	\includegraphics[width=\columnwidth]{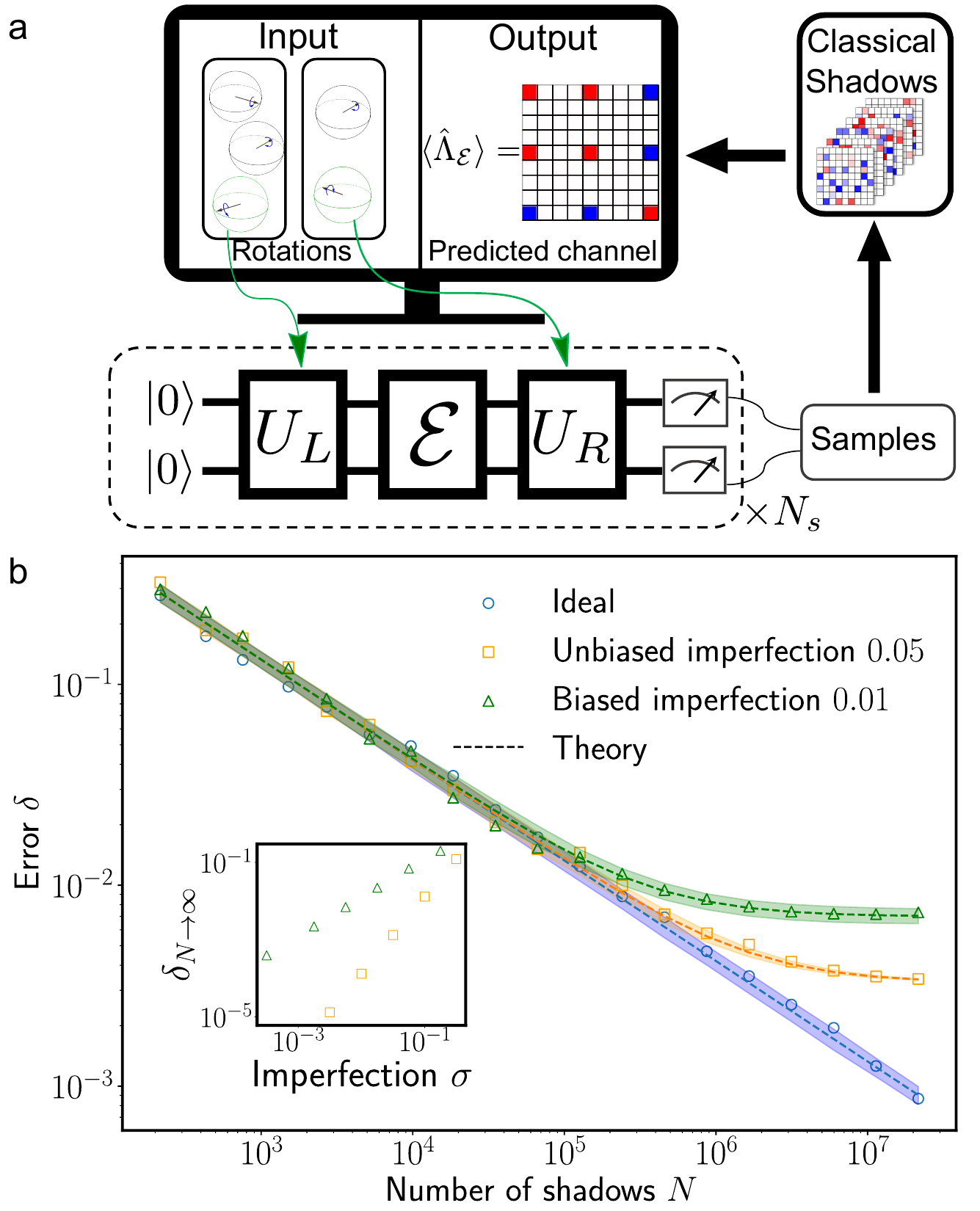}
	\caption{(a)
Schematic representation of the quantum process shadow tomography protocol.
See Algorithm I for the step-by-step description of the protocol.
 (b) 
Scaling of the error $\delta$ as a function of the number of measurements for an optimal choice of $S_L$ and $S_R$ under various assumptions: ideal unitary (blue circles) or in the presence of random imperfections in $U_L$ and $U_R$ that are unbiased (yellow squares) or biased (green triangles). 
The markers represent numerical simulation results for $\mathcal{E}=\mathrm{CZ}$, compared against the analytic theory predictions (dashed lines).
 Inset: the error in the limit of infinite measurement samples does not vanish and depends on the strength of the imperfections.
}
    \label{f4}
\end{figure}

\textit{Quantum Process Tomography.}
Having introduced methods to engineer quantum gates and effective spin Hamiltonian, we turn to describing a practical approach to characterize quantum dynamics in laboratory and to verify our predictions.
More specifically, our approach will be based on the classical shadow process tomography, which is further improved for our purposes~\cite{sqpt_luo,kunjummen2021shadow,helsen2021estimating,elben2022randomized,huang2020predicting}. 
Due to the symmetry of exchanging two qubits, it suffices to identify the unitary $\mathcal{E} \in \mathrm{SU}(3)$ acting on the triplet subspace spanned by $\{\ket{b=-1} = \ket{11}, \ket{b=0} = (\ket{01}+\ket{10})/\sqrt{2}$, $\ket{b = 1} = \ket{00}\}.$  
Our protocol consists of three steps: (i) initializing states by an $\mathrm{SO}(3)$ rotation $U_L$, (ii) evolving them under $\mathcal{E}$, and (iii) performing measurements after applying the second rotation $U_R$.
Our algorithm returns an estimation $\hat{\Lambda}_\mathcal{E}$ of a Choi matrix representation $\Lambda_\mathcal{E}$~\cite{CHOI1975285}, encoded as a $9\times 9$ matrix, of a quantum channel $\mathcal{E}$ (see~\cite{sup} for details). See Figure~\ref{f4} and Algorithm I.

\begin{table}[h!]
    \centering
    \begin{tabularx}{\linewidth}{ccX}
    \hline
     \multicolumn{3}{X}{Algorithm I: Quantum process shadow tomography}\\
    \hline
        1.& \multicolumn{2}{X}{For every pair of $U_L^i \in S_L $ and $U_R^j \in S_R$, repeat the following steps $N_s$ times:}\\
        & a. & Prepare the initial state $\ket{00}$.\\
        & b. & Apply $U_L^i$ to the state, and then apply the unknown quantum channel $\mathcal{E}$ followed by $U_R^j$.\\
        & c. & Measure the qubits and convert the measurement result to $\ket{b}$, where $b=0$ for either $\ket{01}$ or $\ket{10}$, $b = -1$ for $\ket{11}$, and $b=1$ for $\ket{00}$.\\
        & d. & Compute the single-shot quantum channel estimator $\hat{\Lambda}_\mathcal{E} (U_L^i, U_R^j, b)$.\\
        2.& \multicolumn{2}{X}{Compute the average of $\hat{\Lambda}_\mathcal{E}$ for all measurements, which estimates the quantum channel in Choi representation.}\\
    \hline
    \end{tabularx}
    \label{algorithm}
\end{table}

The performance of this algorithm depends on the choice of the sets $S_L$ and $S_R$. Here, we find an optimal choice, namely the smallest subsets of $\mathrm{SO(3)}$ such that the average estimator converges the fastest as the number of measurements increases.
Specifically, define the error between the real channel and the predicted channel as $\delta = ||\langle\hat{\Lambda}_\mathcal{E}\rangle - \Lambda_\mathcal{E}||_\mathrm{F}/9$, where $||\cdot||_\mathrm{F}$ is the matrix Frobenius norm. Then for any choice of rotations, the number of measurements needed for a finite distance $\delta$ scales as  
$N = C(S_L, S_R)(\mathcal{E})/\langle\delta^2\rangle$ [Figure~\ref{f4}(b)], due to statistical errors associated with random measurement outcomes. The function $C(S_L, S_R)(\mathcal{E})$, which characterizes the sample complexity, is in general $\mathcal{E}$-dependent.
We evaluate an upper bound $A(S_L, S_R)$ over all possible $\mathcal{E}$ and find the optimal $S_L$ and $S_R$ by minimizing $A$.

The minimization of $A$ consists of a continuous optimization under given set sizes $|S_L|$ and $|S_R|$ using the gradient descent algorithm, and  a search for the minimum over different set sizes.
In the gradient descent stage, the rotation unitaries $U_L^i$ and $U_R^j$ are parametrized by the rotation axis orientations and rotation angles. We observe that $A$ has a local minimum when $|S_L|=12$ and $|S_R|=9$~\cite{sup}.

In order to model realistic situations, we also analyze the effects of imperfections in experimental control. 
If there are unbiased random imperfections $\sigma$ in the rotation parameters, which can be considered as an effective noisy channel, then there will be a nonzero error $\delta\propto \sigma^2$ even for an infinite number of measurements. 
If the imperfection $\sigma$ in the rotations is systematic and biased, 
then the error in the predicted channel is linear in $\sigma$. 
See Figure~\ref{f4}(b).

\textit{Conclusion.} The scalability, the robustness, and the long coherence time of the fermion pair register make it a promising quantum platform in the near future. In this work, by utilizing the capability of adjusting fermion hopping and fermion-fermion interaction via Feshbach resonance, we show that the fermion pair register is able to realize a universal gate set and quantum many-body simulation. In particular, we develop schemes for the SWAP gate and the controlled-phase gates, as well as the transverse field quantum Ising model simulations. Furthermore, we propose an experimentally efficient shadow process tomography protocol to calibrate quantum gates and probe quantum dynamics in the device. 

Our work brings several implications.
First, our work shows that fermion pair registers are in principle capable of universal quantum computation, explicitly showcasing the computational power of the fermionic quantum gas microscopes within reachable technology.
Second, this work provides a number of new opportunities for utilizing fermion pair registers for practical applications. It will be interesting to prepare entangled resource states based on the present schemes and realize measurement-based quantum computation. The fermion pair registers can be arranged in different 1D and 2D geometries, allowing the study of non-equilibrium dynamics of quantum many-body systems as well as exotic quantum magnetism.

\textit{Acknowledgement.} 
The authors acknowledge Thomas Hartke, Botond Oreg, Carter Turnbaugh, Ningyuan Jia, Martin Zwierlein, and Daniel Mark for insightful discussion and their technical comments. 
XS acknowledges support from MIT UROP and the Lord Foundation. 
DL acknowledges support from the NSF AI Institute for Artificial Intelligence and Fundamental Interactions (IAIFI). This material is based upon work supported by the U.S. Department of Energy, Office of Science, National Quantum Information Science Research Centers, Co-design Center for Quantum Advantage (C2QA) under contract number DE-SC0012704.
SC acknowledges support from the NSF.
\bibliography{reference}

\appendix

\clearpage

\onecolumngrid
\begin{center}
	\noindent\textbf{Supplementary Material}
	\bigskip
		
	\noindent\textbf{\large{}}
\end{center}

\renewcommand\thefigure{A\arabic{figure}}  
\renewcommand\thetable{A\arabic{table}}  
\setcounter{figure}{0}  
\setcounter{table}{0}
\setcounter{secnumdepth}{6}

\section{Detailed Derivation of the Effective Hamiltonian}
Here we provide the details for constructing the effective Hamiltonian Eq.~(1) in the main text. 
We start with analyzing two fermions in a single well.
As explained in Ref.~\cite{Hartke_2022}, we encode a qubit by choosing a pair of particle configurations, whose energy difference is insensitive to the optical lattice depth, therefore the laser intensity.  Next we proceed to consider two-qubit interactions by first constructing the two-qubit basis in the second quantization picture. We introduce the tunable on-site interaction $U$ and the tunneling $J$ between two wells and describe the system using the Fermi-Hubbard model. We explicitly compute all relevant matrix elements and apply the secular approximation to obtain the effective Hamiltonian up to the second order perturbation. Based on the above analysis, we obtain an effective interaction between two qubits.

\subsection{Description of Two Fermions in a Single Well of an Optical Lattice}
Consider the Hamiltonian describing two distinguishable particles in one well in an 1D optical lattice, where we assume that particles are in the ground state of the tightly confined potential in $x$ and $y$ directions:
	\begin{equation}
		H = -\dfrac{\hbar^2 }{2m} \pdv{^2}{z_1^2}- \dfrac{\hbar^2 }{2m} \pdv{^2}{z_2^2}+ V(z_1) + V(z_2) + U a_z \delta(z_1-z_2)
	\end{equation}
where $V(z)$ is the optical lattice potential in the $z$ direction,
	\begin{equation}
 \begin{split}
		V(z) &= VE_R\sin^2\left(\dfrac{\pi z}{a_z}\right)\\
  &= VE_R \left[\left(\dfrac{\pi z}{a_z}\right)^2 - \dfrac{1}{3} \left(\dfrac{\pi z}{a_z}\right)^4 + \dfrac{2}{45} \left(\dfrac{\pi z}{a_z}\right)^6 +O(z^8) \right] = \half m \omega^2 z^2 + \delta V(z),
\end{split}
	\end{equation}
and $E_R = \dfrac{\pi^2 \hbar^2}{2 m a_z^2}$ is the recoil energy.

To the lowest order, the eigenstates of non-interacting particles are the eigenstates of simple harmonic oscillators, denoted by the $\psi_{iS}(z)$, where $i\in \mathbb{N}$ enumerates different energy levels and the index $S \in \{L, R\}$ is introduced for later purposes when we consider two neighboring sites: left ($L$) and right ($R$). The wavefunctions $\psi_{i,S}(z)$ for different $S$ are related to each other via spatial translation.
In the absence of interaction $U$, we seek for a pair of nearly degenerate antisymmetric wavefunctions satisfying the fermionic statistics.
In this work, we only consider two fermions forming a spin singlet, hence we focus on constructing a pair of spatially symmetric wavefunctions with two energy quanta, $\ket{1_S,1_S}$ and $\ket{0_S,2_S}_{sym}$. To the zeroth order, they are
\begin{equation}
\begin{split}
    \braket{z_1,z_2}{1_S,1_S}&\equiv\psi_{1S}(z_1)\psi_{1S}(z_2),\\
    \braket{z_1,z_2}{0_S,2_S}_{sym}&\equiv\dfrac{1}{2}(\psi_{0S}(z_1)\psi_{2S}(z_2) + \psi_{2S}(z_1)\psi_{0S}(z_2)).
\end{split}
\end{equation}
Then, the qubit states are defined as $\ket{0} = \ket{1_S,1_S} \otimes (\ket{\up\down}-\ket{\down\up})/\sqrt{2}$ and $\ket{1} = \ket{0_S,2_S}_{sym} \otimes (\ket{\up\down}-\ket{\down\up})/\sqrt{2}$. Due to the anharmonicity $\delta V$, the energy of these two states are not equal, where the first order energy corrections are:
\begin{equation}
\begin{split}
    \bra{0}(\delta V(z_1)+\delta V(z_2))\ket{0}&=-\dfrac{5}{2}E_R + O(V^{-\frac{1}{2}})E_R,\\
    \bra{1}(\delta V(z_1)+\delta V(z_2))\ket{1}&=-\dfrac{7}{2}E_R + O(V^{-\frac{1}{2}})E_R.
\end{split}
\end{equation}
From Figure~1 in the main text (as well as Ref.~\cite{Hartke_2022}) one can see that the interaction $\hat{U}$ is diagonalized in the $\{\ket{+}, \ket{-}\}$ basis. Hence in the $\{\ket{0}, \ket{1}\}$ basis, the interaction $\hat{U}$ contains off-diagonal terms.

\subsection{Fermi-Hubbard Hamiltonian Description}
Using the second quantization description, an $m$-particle state can be expanded in a Fock space basis $\ket{n_1,\cdots, n_{k} ,\cdots,n_m}$ where each $n_k$ is the occupation number of the state $\psi_k$. Here, $k$ is a composite label of the orbital, $k = i_S \sigma$, where $S \in \{L, R\}$ denotes one of the two wells, $i \in \{0, 1, 2, \cdots\}$ is the energy level, and $\sigma \in \{\up, \down\}$ is the spin polarization. To simplify the notation, we use the composite label to represent the occupation of a state. Therefore, the two-qubit basis can be represented as
\begin{align}
    \ket{00} &= \ket{{1_L\uparrow} {1_L\downarrow} {1_R\uparrow} {1_R\downarrow}} \\
    \ket{01} &= \frac{1}{\sqrt{2}} (\ket{{1_L\uparrow} {1_L\downarrow} {0_R\uparrow} {2_R\downarrow}} - \ket{{1_L\uparrow} {1_L\downarrow} {0_R\downarrow} {2_R\uparrow}}) \\
    \ket{10} &= \frac{1}{\sqrt{2}} (\ket{{0_L\uparrow} {2_L\downarrow} {1_R\uparrow} {1_R\downarrow}} - \ket{{0_L\downarrow} {2_L\uparrow} {1_R\uparrow} {1_R\downarrow}}) \\
    \ket{11} &=        
    \frac{1}{2} (\ket{{0_L\up}{2_L\down}{0_R\up}{2_R\down}}-\ket{{0_L\down}{2_L\up}{0_R\up}{2_R\down}}
    -\ket{{0_L\up}{2_L\down}{0_R\down}{2_R\up}}+\ket{{0_L\down}{2_L\up}{0_R\down}{2_R\up}}).
\end{align}
We use $c_{iS\sigma}$ and $c^\dagger_{iS\sigma}$ to denote the annihilation and the creation operator of a particle in state $k=i_S\sigma$.

We consider that the dynamics of fermions in the trap is described by the Fermi-Hubbard model
$H = H_E + H_U + H_J$, where each term describes the orbital energy, the on-site interaction, and the nearest-neighbor tunneling, respectively. Specifically, the orbital energy term $H_E = \sum_{i, S, \sigma}E_i c_{iS\sigma}^\dagger c_{iS\sigma}$. The on-site interaction is
\begin{equation}
    H_U = \sum_{S, i, j, k, l, \sigma_1, \sigma_2}U_{ijkl}c_{iS\sigma_1}^\dagger c_{ jS\sigma_2}^\dagger c_{kS\sigma_1} c_{lS\sigma_2}
\end{equation}
where $U_{ijkl}$ are determined by the $s$-wave Feshbach resonance.
The coefficients $U_{ijkl}$ are given by
\begin{equation}\label{eq:U}
    U_{ijkl} = \frac{1}{2}\int \dd z_1 \dd z_2 \psi_i^*(z_1)\psi_j^*(z_2) \hat{U} \psi_k(z_2)\psi_l(z_1).
\end{equation}

The tunneling term $H_J = - \sum_{i,j, \sigma} J_{ij}(c_{iL \sigma}^\dagger c_{jR \sigma} + c_{jR \sigma}^\dagger c_{iL\sigma})$ describes the single particle hopping between adjacent sites, and the hopping amplitude does not depend on the energy level $i$ because the overlap of all relevant orbitals, which are ground states in the $xy$-plane, is independent of $z$-orbitals. The hopping process between different energy levels can be neglected under secular approximation when $\omega_0 \gg J_{ij}$.

By the Fermionic statistics and the fact that $\hat{U}$ conserves spin, the interaction is nonzero only if two Fermions have opposite spins and the final state has the same total spin as the initial state. In addition,  in the deep potential limit, the interaction is on-site because the interaction between particles in different sites is strongly suppressed due to exponentially decaying spatial wavefunctions. As an example, we explicitly compute the following terms:
\begin{equation}
    U_{0220} = U_{1111} = \frac{3}{16}\sqrt{\frac{\pi}{2}}U,\ U_{0211} = \frac{\sqrt{\pi}}{16}U.
\end{equation}
Thus, the energies of both $\ket{0}$ and $\ket{1}$ are $4U_{0220}=4U_{1111}$. The off-diagonal matrix elements mentioned before are $\bra{0}\hat{U}\ket{1} = \bra{1}\hat{U}\ket{0} = \sqrt{2}U_{0211}$. 

To use the secular approximation, we set $\tilde{\ket{\psi}} = e^{i H_E t} \ket{\psi}$. The Hamiltonian in the interaction picture is $\tilde{H} = \tilde{H_U} + \tilde{H_J}$, where
\begin{align}
    \tilde{H_U} = \langle e^{iH_Et}H_Ue^{-iH_Et}\rangle_t,\\
    \tilde{H_J} = \langle e^{iH_Et}H_Je^{-iH_Et}\rangle_t.
\end{align}

\subsection{Derivation of the Effective Hamiltonian Restricted to the Qubit Hilbert Space}
For the discussion of CPHASE gate, we assume $J \ll U \ll E_R$. We first apply the secular approximation to the full Hamiltonian $H = H_E + H_U + H_J$ to eliminate $H^{(0)} = H_E$ and obtain $\tilde{H} = \tilde{H_U} + \tilde{H_J}$ by moving into an interaction picture. Since the hopping amplitude only depends on the overlap in $xy$-plane, it is independent of $z$-orbitals. Off-resonant tunneling and long-range tunneling are suppressed in our gate engineering protocol. Hence, we assume $J_{ij} = J \delta_{ij}$. We note that our Hamiltonian has the symmetry of exchanging two qubits, therefore the interaction $U$ do not lift the degeneracy between dressed versions of $\ket{01}$ and $\ket{10}$.  Furthermore, the particle configurations of the above two degenerate states require at least four hoppings to couple each other, so that the second order process with $J$ (involving only two hoppings) allows us to perform non-degenerate perturbation theory to determine energy shifts. To compute the perturbative corrections, we compute the matrix elements between any state above with other eigenstates of the Hamiltonian $\tilde{H_U}$ (which should be $H_E+H_U$, but we are ignoring the state dressing in the secular approximation, which leads to additional correction factor $O(U/E_R)$ relative to the second order effect in $J/U$). It is obvious that $\tilde{H_J}\ket{00}=0$ due to the Pauli exclusion principle, hence $\bra{f}\tilde{H_J}\ket{00} = 0$. The following matrix entries contribute to the second order perturbation:
\begin{align}
    &\dfrac{|\bra{{0_L\up}{0_L\down}{2_L\up}{2_R\down}}\tilde{H_J}\ket{11}|^2}{E_{\ket{{0_L\up}{0_L\down}{2_L\up}{2_R\down}}} - E_{11}} = \dfrac{\left(-\frac{1}{2}J\right)^2}{\left(\frac{11}{8}\sqrt{\frac{\pi}{2}}-\frac{3}{2}\sqrt{\frac{\pi}{2}}\right)U}\\
    &\dfrac{|\bra{{0_L\down}{2_L\up}{2_L\down}{0_R\up}}\tilde{H_J}\ket{11}|^2}{E_{\ket{{0_L\down}{2_L\up}{2_L\down}{0_R\up}}} - E_{11}} = \dfrac{\left(\frac{1}{2}J\right)^2}{\left(\frac{65}{64}\sqrt{\frac{\pi}{2}}-\frac{3}{2}\sqrt{\frac{\pi}{2}}\right)U}\\
    &\dfrac{|\bra{{2_L\down}{0_R\up}{1_R\up}{1_R\down}}\tilde{H_J}\ket{10}|^2}{E_{\ket{{2_L\down}{0_R\up}{1_R\up}{1_R\down}}} - E_{10}} = \dfrac{\left(-\frac{1}{\sqrt{2}}J\right)^2}{\left(\frac{5}{4}\sqrt{\frac{\pi}{2}}-\frac{3}{2}\sqrt{\frac{\pi}{2}}\right)U}\\
    &\dfrac{|\bra{{0_L\up}{1_R\up}{1_R\down}{2_R\down}}\tilde{H_J}\ket{10}|^2}{E_{\ket{{0_L\up}{1_R\up}{1_R\down}{2_R\down}}} - E_{10}} = \dfrac{\left(\frac{1}{\sqrt{2}}J\right)^2}{\left(\frac{19}{16}\sqrt{\frac{\pi}{2}}-\frac{3}{2}\sqrt{\frac{\pi}{2}}\right)U}
\end{align}
and denote 
\begin{align}
    \ket{\psi_a} &= c_1 \ket{{0_L\up}{1_L\up}{2_L\down}{1_R\down}} + c_2 \ket{{0_L\down}{1_L\up}{2_L\up}{1_R\down}} + c_3 \ket{{0_L\up}{1_L\down}{2_L\up}{1_R\down}}\\
    \ket{\psi_b} &= c_3 \ket{{0_L\up}{1_L\up}{2_L\down}{1_R\down}} + c_2 \ket{{0_L\down}{1_L\up}{2_L\up}{1_R\down}} + c_1 \ket{{0_L\up}{1_L\down}{2_L\up}{1_R\down}}\\
    \ket{\psi_c} &= \frac{1}{\sqrt{3}} (\ket{{0_L\up}{1_L\up}{2_L\down}{1_R\down}} +  \ket{{0_L\down}{1_L\up}{2_L\up}{1_R\down}} + \ket{{0_L\up}{1_L\down}{2_L\up}{1_R\down}})
\end{align}
where $c_1 = \frac{-3+\sqrt{3}}{6}$, $c_2 = -\frac{1}{\sqrt{3}}$, $c_3 = \frac{3+\sqrt{3}}{6}$. These three states are eigenstates of the Hamiltonian with eigenvalues $\frac{21\pm\sqrt{3}}{16}\sqrt{\frac{\pi}{2}}U$ and 0, respectively, and the first two of them have non-vanishing matrix entries with the logical states $\ket{10}$.

\begin{align}
    & \dfrac{|\bra{\psi_a}\tilde{H_J}\ket{10}|^2}{E_{\ket{\psi_a}}-E_{10}} = \dfrac{(c_1-c_2)^2\left(\frac{1}{\sqrt{2}}J\right)^2}{\left(\frac{21+\sqrt{3}}{16}\sqrt{\frac{\pi}{2}}-\frac{3}{2}\sqrt{\frac{\pi}{2}}\right)U}\\
    & \dfrac{|\bra{\psi_b}\tilde{H_J}\ket{10}|^2}{E_{\ket{\psi_b}}-E_{10}} = \dfrac{(c_3-c_2)^2\left(\frac{1}{\sqrt{2}}J\right)^2}{\left(\frac{21-\sqrt{3}}{16}\sqrt{\frac{\pi}{2}}-\frac{3}{2}\sqrt{\frac{\pi}{2}}\right)U} \\
    & \dfrac{|\bra{\psi_c}\tilde{H_J}\ket{10}|^2}{E_{\ket{\psi_c}}-E_{10}} = 0
\end{align}

Due to the symmetry between $\ket{01}$ and $\ket{10}$, the energy shift of $\ket{01}$ equals that of $\ket{10}$. Meanwhile, the energy of $\{\tilde{\ket{00}},\tilde{\ket{01}},\tilde{\ket{10}},\tilde{\ket{11}}\}$ in the effective Hamiltonian of Eq.~(1) in the main text have the energies $\omega-g/2, g/2,g/2,-\omega-g/2$ up to a constant shift. Matching with the second order corrections computed above, we can derive
\begin{equation}\label{eq:g}
    g = \frac{3152}{155}\sqrt{\frac{2}{\pi}}\frac{J^2}{U},
\end{equation}
and 
\begin{equation}\label{eq:omega}
    \omega = \dfrac{\pi U^2}{16E_R} - \dfrac{156}{31} \sqrt{\frac{2}{\pi}}\frac{J^2}{U}.
\end{equation}

\section{Adiabatic Protocol for Controlled-phase Gate}
We define the gate fidelity $f$ of a protocol as the fidelity between the theory-predicted target state and the state evolved under $H$, averaged over Haar random initial states
\begin{equation}\label{eq:fid}
    f = \mathbb{E}_{\ket{\psi}\sim \mathrm{Haar\ random}} |\bra{\psi} P^\dagger U_{\mathrm{theory}}^\dagger P e^{-\mathrm{i}\int \dd t H(t)}\ket{\psi}|^2,
\end{equation}
where $U_{\mathrm{theory}}$ is the target gate, $P$ is an isometry between the logical subspace $\mathcal{H}_L^{\otimes 2}$ and the full Hilbert space of four fermionic particles involving all relevant vibrational excited states. To maximize the fidelity of the gate, the hopping amplitude $J$ is adiabatically tuned such that a system initialized to any eigenstate $\ket{\psi}$ of $H_E+H_U$ is expected to remain being an eigenstate $\tilde{\ket{\psi}}$ of $H(t)$ at any given time $t$ (Figure 3 in the main text). Empirically, the function $J(t)$ with a continuous first derivative leads to a high fidelity. 
Hence, we utilize the following function~\cite{conlon2019error,coopmans2021protocol}:
\begin{equation}\label{eq:J}
    J(t) = \left\{ 
    \begin{aligned}
        &J_0 \sin^2 (\pi t/(2\eta T)), &0\le t < \eta T,\\
        &J_0, &\eta T \le t < (1-\eta) T, \\
        &J_0 \sin^2 (\pi/2 (1- (t-(1-\eta) T)/(\eta T))), & (1-\eta) T \le t < T, \\
        &0, &\mathrm{otherwise}
    \end{aligned}
    \right.
\end{equation}
where $\eta\in(0, 0.5]$ is the portion of time for adiabatic ramping. The induced phase accumulated throughout this protocol is 
\begin{equation}
    \varphi(T) =\int_{0}^{T} \dd t\ g(t) = \left(1-\dfrac{5}{4} \eta\right)gT.
\end{equation}
Qualitatively, larger $\eta$ stands for slower ramping and slower gate speed. In our gate simulation, we choose $\eta = 0.3$.

Suppose that a CPHASE gate has time duration $T$, then the fourth order effect in $J/U$ perturbs the state on the order of $O(J_0^4 T/U_0^3)$, i.e. $\ket{\psi} = (1 - O(J_0^8 T^2/U_0^6))\ket{\psi_\mathrm{theory}} + O(J_0^4 T/U_0^3) \ket{\psi_\perp}$. The infidelity $1-f = 1-|\bra{\psi_\mathrm{theory}}\ket{\psi}|^2$ is proportional to $J_0^8 T^2/U_0^6$. For a certain CPHASE gate of induced phase $\varphi$, the time duration $T$ is proportional to $\varphi U_0/J_0^2$, so the infidelity of a CPHASE gate is proportional to $(J_0/U_0)^4$ (Figure 3(b) in the main text). Meanwhile, the secular approximation leads to a correction $O(J_0^2 T/E_R) \ket{\psi_\perp}$ of the state, resulting in an infidelity proportional to $(U_0/E_R)^2$ for the same reason.

We use numerical simulation to verify the effective Hamiltonians in Eq.~(1) of the main text and the gate protocols. Our numerical simulation in Figure~3 is based on exact diagonalization of the Fermi-Hubbard model Hamiltonian. Since all dynamics preserves total particle number, energy quanta, and spin, the subspace with 4 particles, 4 energy quanta, and zero total spin is an invariant subspace of the Hamiltonian. Therefore, the numerical simulation can be restricted to the subspace with dimensions being significantly reduced (if we truncate the harmonic oscillator energy ladders containing three lowest-energy states, the dimension of the subspace, denoted by $\mathcal{H}_{inv}$, is 59). For a time-dependent Hamiltonian, we calculate the discrete-time evolution of the system, where the time step $\delta$ is chosen such that the discretization error is small, i.e. $\frac{\dd \langle H\rangle}{\dd t} \delta^2 \ll 1$, and in each step the Hamiltonian is diagonalized to calculate the unitary evolution matrix $e^{-i H \delta}$. The value of $\delta$ is sufficiently small such that the simulation error is negligible comparing with gate infidelity.

To compute the fidelity of gate protocols, we simulate the system with the full Hamiltonian in the invariant subspace and compare with the simulation using the predicted effective Hamiltonian in a smaller subspace, $\mathcal{H}_L^{\otimes 2}$. To capture how well the protocols keep final states in $\mathcal{H}_L^{\otimes 2}$, in Eq.~\ref{eq:fid} we first project the final state to the logical subspace by the isometry $P: \mathcal{H}_{inv} \rightarrow \mathcal{H}_L^{\otimes 2}$ and then compute the fidelity. 

For gate protocols, we compute the time evolution of four basis states, and then compute the fidelity in Eq.~\ref{eq:fid} by randomly sampling initial states with respect to Haar measure and then average over 300000 samples.

\section{Derivation and Numerical Simulation of the Quantum Ising Model}
We derive the effective Ising model Hamiltonian (Eq.~(2) in the main text) and numerically verify our prediction.
For the following Hamiltonian describing a single qubit in a single well,
\begin{equation}
    H_{\mathrm{1-qubit}} = \begin{bmatrix}
        \dfrac{E_R}{2} & \sqrt{\dfrac{\pi}{128}}(U_0 + U_1 \cos((E_R + \Delta) t)) \\
        \sqrt{\dfrac{\pi}{128}}(U_0 + U_1 \cos((E_R + \Delta) t)) & -\dfrac{E_R}{2}
    \end{bmatrix},
\end{equation}
when $E_R \gg U_0, U_1$, and $\Delta \ll E_R$, in a frame rotating with angular frequency $E_R + \Delta$, $\tilde{\ket{\psi}} = e^{i (E_R+\Delta) \sigma_z t/2} \ket{\psi}$, the effective Hamiltonian can be approximated as
\begin{equation}
    H_\mathrm{1-qubit,\, eff} = \langle e^{iH^{(0)}t}(H-H^{(0)})e^{-iH^{(0)}t}\rangle_t = \begin{bmatrix}
        -\dfrac{\Delta}{2} & \half\sqrt{\dfrac{\pi}{128}} U_1 \\
        \half\sqrt{\dfrac{\pi}{128}}U_1 & \dfrac{\Delta}{2}
    \end{bmatrix}
\end{equation}
where $H^{(0)} = (E_R + \Delta) \sigma_z /2$. This introduces the Rabi oscillation in Eq.~(2).

We numerically simulate the Ising dynamics between two sites to verify the effective Hamiltonian Eq.~(2) in the main text using the exact diagonalization method, described in the previous section. Similar to the gate protocol, the hopping term $J$ is turned on adiabatically, and the modulation $U_1$ is turned on instantly afterwards. We also simulate in the logical space $\mathcal{H}_L^{\otimes 2}$ with the Ising Hamiltonian 
\begin{equation}
H = \mathbf{c}(t) \cdot (\boldsymbol{\sigma}^1 + \boldsymbol{\sigma}^2) + g(t) \sigma_z^1 \sigma_z^2
\end{equation}
where $\mathbf{c}(t) = (c_x(t), c_y(t), c_z(t))$. The function forms of $\mathbf{c}(t)$ and $g(t)$ are chosen according to Eq.~(2) in the main text, Eq.~\ref{eq:g}, \ref{eq:omega}, and \ref{eq:J}, but we treat the numerical prefactors as variables. For example, $c_z(t) = \alpha J(t)^2/U_0 + \beta$ where $\alpha$ and $\beta$ are free parameters. We then maximize the fidelity in Eq.~\ref{eq:fid} to obtain the optimal parameters in $\mathbf{c}(t)$ and $g(t)$.

\begin{figure}[h!]
    \centering
    \includegraphics[width=0.53\linewidth]{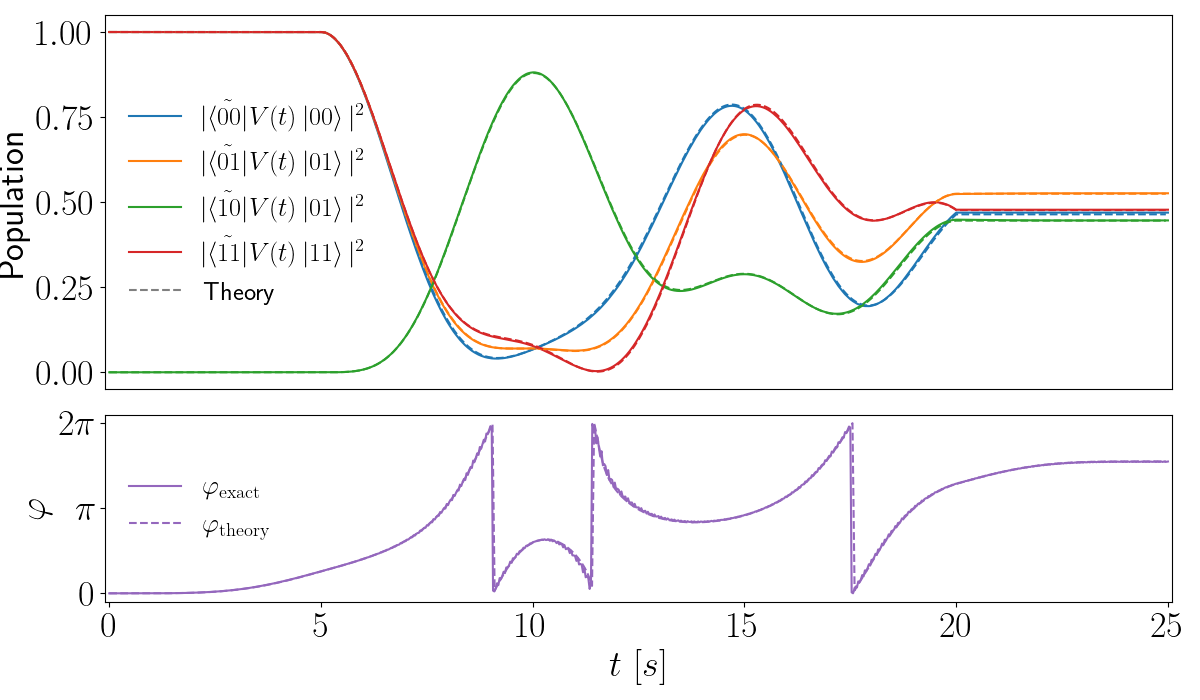}    \includegraphics[width=0.44\linewidth]{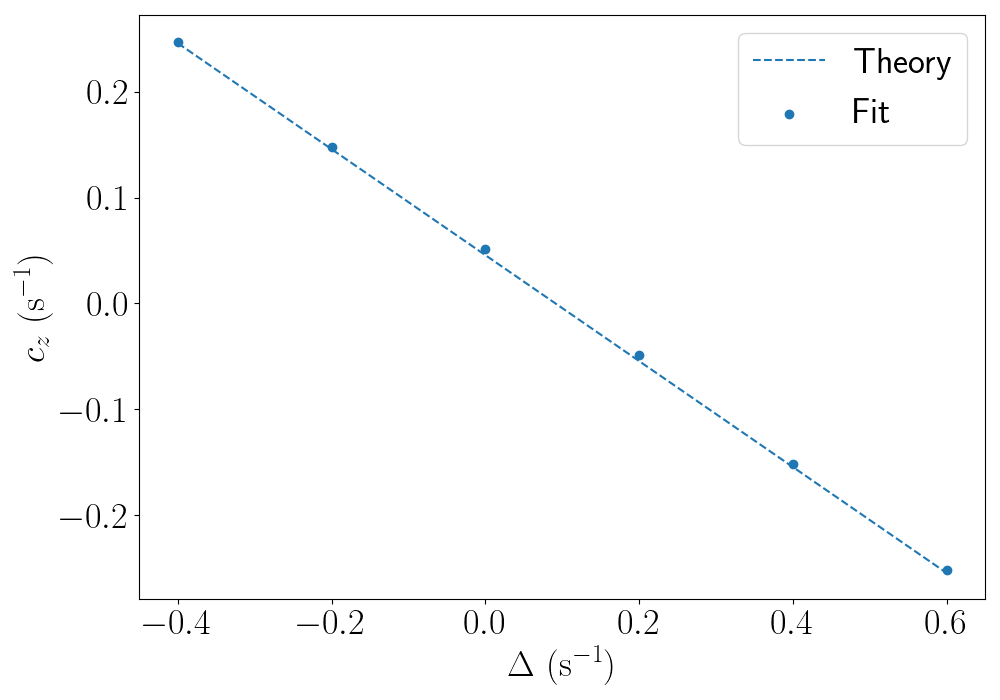}
    \caption{Left: Ising model simulation: $E_R = 2\pi \cdot 140.76 \mathrm{Hz}, U_0 = 30 \mathrm{s}^{-1}, U_1 = 2\mathrm{s}^{-1}, J = 0.9 \mathrm{s}^{-1}$, fidelity$=0.999699$. Right: The fit $c_z$ (during the period when both $J$ and $U_1$ are constant) versus various detunings $\Delta$.}
    \label{fig:ising-1}
\end{figure}

Indeed, Figure~\ref{fig:ising-1} Left shows promising agreement between the theory and the numerical simulation, and by choosing various detunings, we verify that the $c_z$ term behaves as Eq.~(2) in the main text and can be tuned to zero by choosing a proper detuning $\Delta$ (Figure~\ref{fig:ising-1} Right).

\section{Quantum Process Tomography}
The goal of quantum process tomography is to predict an unknown quantum dynamics, where in this paper we consider the unitary quantum dynamics that respect qubit exchange symmetry, with as few measurements as possible. Therefore, with initial state $\ket{00}$, the triplet subspace $\mathrm{span}\{\ket{00}, (\ket{01}+\ket{10})/\sqrt{2}, \ket{11}\}$ is an invariant subspace of the relevant quantum dynamics, and the basis vectors are denoted by $\ket{b},\ b \in \{1, 0, -1\}$. Per experimental limitation and convenience, we design a protocol requiring only global manipulations and qubit readout in conventional logical basis. Any global rotation can be represented in the $\ket{b}$ basis in the following way:
\begin{equation}\label{eq:rotations}
    U(\mathbf{n(\theta, \phi)}, \alpha) = e^{-i \alpha\, \mathbf{n}\cdot \mathbf{J}}
\end{equation}
where $\mathbf{n(\theta, \phi)} = (\sin\theta \cos\phi, \sin\theta\sin\phi, \cos\theta)$, and 
$$J_x = \frac{1}{\sqrt{2}}\begin{bmatrix}
    0 & 1 & 0\\
    1 & 0 & 1\\
    0 & 1 & 0
\end{bmatrix},\ J_y = \frac{1}{\sqrt{2}}\begin{bmatrix}
    0 & -i & 0\\
    i & 0 & -i\\
    0 & i & 0
\end{bmatrix},\ J_z = \begin{bmatrix}
    1 & 0 & 0\\
    0 & 0 & 0\\
    0 & 0 & -1
\end{bmatrix}.$$

The following representation, the Choi representation, encodes full information of a quantum channel $\mathcal{E}$ (\cite{CHOI1975285}):
\begin{equation}
    \Lambda_\mathcal{E} = \sum_{a, b = -1}^{1} \ketbra{a}{b} \otimes \mathcal{E}(\ketbra{a}{b}).
\end{equation}
The reconstructive formula in the algorithm is 
\begin{equation}\label{eq:reconstruction}
    \hat{\Lambda}_\mathcal{E}(U_L^i, U_R^j, b)=M_L^{-1}(U_L^i \ketbra{1}{1}U_L^{i\dagger})^T \otimes M_R^{-1}(U_R^{j\dagger} \ketbra{b}{b}U_R^j) ,
\end{equation}
where
\begin{equation}\label{eq:channels}
\begin{aligned}
    &M_L(\rho)=\sum_{U_L^i\in S_L} \frac{\mathrm{Tr}((U_L^i\ketbra{1}{1}U_L^{i\dagger})^T \rho) (U_L^i \ketbra{1}{1} U_L^{i\dagger})^T}{|S_L|},\\
    &M_R(\rho)=\sum_{U_R^j\in S_R}\sum_{b=-1}^{1} \frac{\mathrm{Tr}(U_R^{j\dagger}\ketbra{b}{b}U_R^j \rho) U_R^{j\dagger} \ketbra{b}{b} U_R^j}{|S_R|}.
\end{aligned}
\end{equation}

The linear maps $M_L$ and $M_R$ depend on the sets $S_L$ and $S_R$, and they can be made invertible since global rotations can determine the channel $\mathcal{E}$. Here we provide the derivation of Eq.~\ref{eq:reconstruction}..

For a matrix $A=(A_{ij,kl})_{i,j=1, \cdots, n}^{k,l=1, \cdots, n}$, denote the partial trace as $(\mathrm{Tr}_1(A))_{jl}=\sum_i A_{ij, il}$ and $(\mathrm{Tr}_2 (A))_{ik}=\sum_j A_{ij, kj}$. Let 
\begin{equation}
(\Lambda_\mathcal{E})_{ij, kl} = \mathcal{E}(\ketbra{i}{k})_{jl}.
\end{equation}
Then we have $\mathcal{E}(\rho)=\mathrm{Tr}_1 ((\rho^T \otimes \mathbb{I}) \Lambda_\mathcal{E})$. 

Consider the circuit in Figure~4 in the main text. We compute the average of $(U_L^i\ketbra{1}U_L^{i\dagger})^T \otimes (U_R^{j\dagger}\ketbra{b}U_R^j)$ over possible measurement outcomes and all $U_L^i \in S_L, U_R^j\in S_R$.
\begin{equation}\label{eq:pf}
\begin{split}
    &\dfrac{1}{|S_L||S_R|}\sum_{U_L^i, U_R^j} \mathbb{E}_b(U_L^i\ketbra{1}U_L^{i\dagger})^T \otimes (U_R^{j\dagger}\ketbra{b}U_R^j)\\
    =&\dfrac{1}{|S_L||S_R|}\sum_{U_L^i, U_R^j, b}(U_L^i\ketbra{1}U_L^{i\dagger})^T \otimes (U_R^{j\dagger}\ketbra{b}U_R^j)\ \mathrm{Tr}(\ketbra{b} U_R^j \mathcal{E}(U_L^i\ketbra{1}U_L^{i\dagger}) U_R^{j\dagger})\\
    =&\dfrac{1}{|S_L||S_R|}\sum_{U_L^i, U_R^j, b}(U_L^i\ketbra{1}U_L^{i\dagger})^T \otimes (U_R^{j\dagger}\ketbra{b}U_R^j)\ \mathrm{Tr}_2(U_R^{j\dagger}\ketbra{b} U_R^j \mathrm{Tr}_1((U_L^i\ketbra{1}U_L^{i\dagger})^T\otimes \mathbb{I})\Lambda_\mathcal{E}))\\
    =&\dfrac{1}{|S_L||S_R|}\sum_{U_L^i, U_R^j, b}(U_L^i\ketbra{1}U_L^{i\dagger})^T \otimes (U_R^{j\dagger}\ketbra{b}U_R^j)\ \mathrm{Tr}((U_L^i\ketbra{1}U_L^{i\dagger})^T\otimes U_R^{j\dagger}\ketbra{b} U_R^j )\Lambda_\mathcal{E})\\
    =& (M_L \otimes M_R)(\Lambda_\mathcal{E})
\end{split}
\end{equation}
by Eq.~\ref{eq:channels}. Hence by applying $M_L^{-1} \otimes M_R^{-1}$ on Eq.~\ref{eq:pf}, we obtain
\begin{equation}
    \Lambda_\mathcal{E} = (M_L^{-1} \otimes M_R^{-1})\dfrac{1}{|S_L||S_R|}\sum_{U_L^i, U_R^j} \mathbb{E}_b(U_L^i\ketbra{1}U_L^{i\dagger})^T \otimes (U_R^{j\dagger}\ketbra{b}U_R^j) = \langle\hat{\Lambda}_\mathcal{E}(U_L^i,U_R^j,b)\rangle.
\end{equation}
where $\hat{\Lambda}_\mathcal{E}(U_L^i, U_R^j, b_{ij})$ is defined in Eq.~\ref{eq:reconstruction}.

The sample complexity, $C(S_L, S_R)(\mathcal{E})$, defined by the number of measurements to achieve certain precision $\delta(N) = \sqrt{C(S_L, S_R)(\mathcal{E})}/\sqrt{N}$, is given by
\begin{equation}
    C(S_L, S_R)(\mathcal{E}) = \dfrac{|S_L||S_R|}{81}\sum_{\mathbf{b}\in\{1, 0, -1\}^{\otimes|S_L||S_R|}} \left(\prod_{i,j}\mathrm{Tr}(U_R^j\mathcal{E}(U_L^i \ketbra{1}{1}U_L^{i\dagger})U_R^{j\dagger} \ketbra{b_{ij}}{b_{ij}})\right)\, \left|\left|\sum_{i,j}\dfrac{\hat{\Lambda}_\mathcal{E}(U_L^i, U_R^j, b_{ij})}{|S_L||S_R|} - \Lambda_\mathcal{E}\right|\right|_\mathrm{F}^2,
\end{equation}
and the quantity
\begin{equation}
    A(S_L,S_R)=\dfrac{1}{81|S_L||S_R|}\sum_{U_L^i,U_R^j}||M_L^{-1} (U_L^i \ketbra{1}{1}U_L^{i\dagger})||_\mathrm{F}^2\, \max\limits_{b} ||M_R^{-1} (U_R^j\ketbra{b}{b} U_R^{j\dagger})||_\mathrm{F}^2
\end{equation}
is an upper bound of this value. To see this, consider the following quantity
\begin{equation}
    B(S_L, S_R)(\mathcal{E}) = \dfrac{1}{81|S_L||S_R|}\sum_{U_L^i, U_R^j, b} \mathrm{Tr}(U_R^j\mathcal{E}(U_L^i \ketbra{1}{1}U_L^{i\dagger})U_R^{j\dagger} \ketbra{b})\ ||\hat{\Lambda}_\mathcal{E}(U_L^i, U_R^j, b) - \Lambda_\mathcal{E}||^2_\mathrm{F}.
\end{equation}
First we show that that $B < A$. Notice that
\begin{equation}\label{eq:42}
    B(S_L, S_R)(\mathcal{E}) < \dfrac{1}{81|S_L||S_R|}\sum_{U_L^i, U_R^j, b} \mathrm{Tr}(U_R^j\mathcal{E}(U_L^i \ketbra{1}{1}U_L^{i\dagger})U_R^{j\dagger} \ketbra{b})\ ||M_L^{-1} (U_L^i \ketbra{1}{1}U_L^{i\dagger})||^2_\mathrm{F}\ ||M_R^{-1} (U_R^j\ketbra{b}{b} U_R^{j\dagger})||^2_\mathrm{F},
\end{equation}
followed from the fact that the variance of a random variable is smaller than its second moment, which are the L.H.S. and the R.H.S. of~\ref{eq:42}. The only $\mathcal{E}$-dependence is in the trace term, which stands for the probability of getting $b$. Hence, by replacing the sum over $b$ with the largest $||M_R^{-1} (U_R^j\ketbra{b}{b} U_R^{j\dagger})||_\mathrm{F}^2$, we get $A(S_L, S_R)$ which is independent of $\mathcal{E}$ and provides an upper bound of $B(S_L, S_R)(\mathcal{E})$.

Hence it suffices to show that $C \le B$. Notice that the trace term in \ref{eq:42} represents the probability of measuring $b$, we will denote this term as $P(b|i,j)$ for simplicity. Then it suffices to prove that
\begin{equation}\label{laststep}
    \sum_{\mathbf{b}\in\{1, 0, -1\}^{\otimes|S_L||S_R|}} \prod_{i, j} P(b_{ij}|i, j) \left|\left|\sum_{i, j} f_{ij}(b_{ij})\right|\right|_\mathrm{F}^2 \le \sum_{\mathbf{b}\in\{1, 0, -1\}^{\otimes|S_L||S_R|}}\prod_{i, j} P(b_{ij}|i, j) \sum_{i, j} \left|\left|f_{ij}(b_{ij})\right|\right|_\mathrm{F}^2 
\end{equation}
where $f_{ij}(b_{ij}) = \hat{\Lambda}_\mathcal{E} (U_L^i, U_R^j, b_{ij}) - \Lambda_\mathcal{E}$ with 
\begin{equation}
\sum_{\mathbf{b}\in\{1, 0, -1\}^{\otimes|S_L||S_R|}} \prod_{i, j} P(b_{ij}|i, j) \sum_{i, j} f_{ij}(b_{ij}) = 0.
\end{equation}
The difference between the L.H.S. and the R.H.S. of \ref{laststep} is $-\sum_{ij} ||\sum_{b} P(b|i,j) f_{ij}(b)||^2_\mathrm{F} \le 0$. Hence $A > B \ge C$.

\begin{figure}[htbp]
    \centering\includegraphics[width=0.5\linewidth]{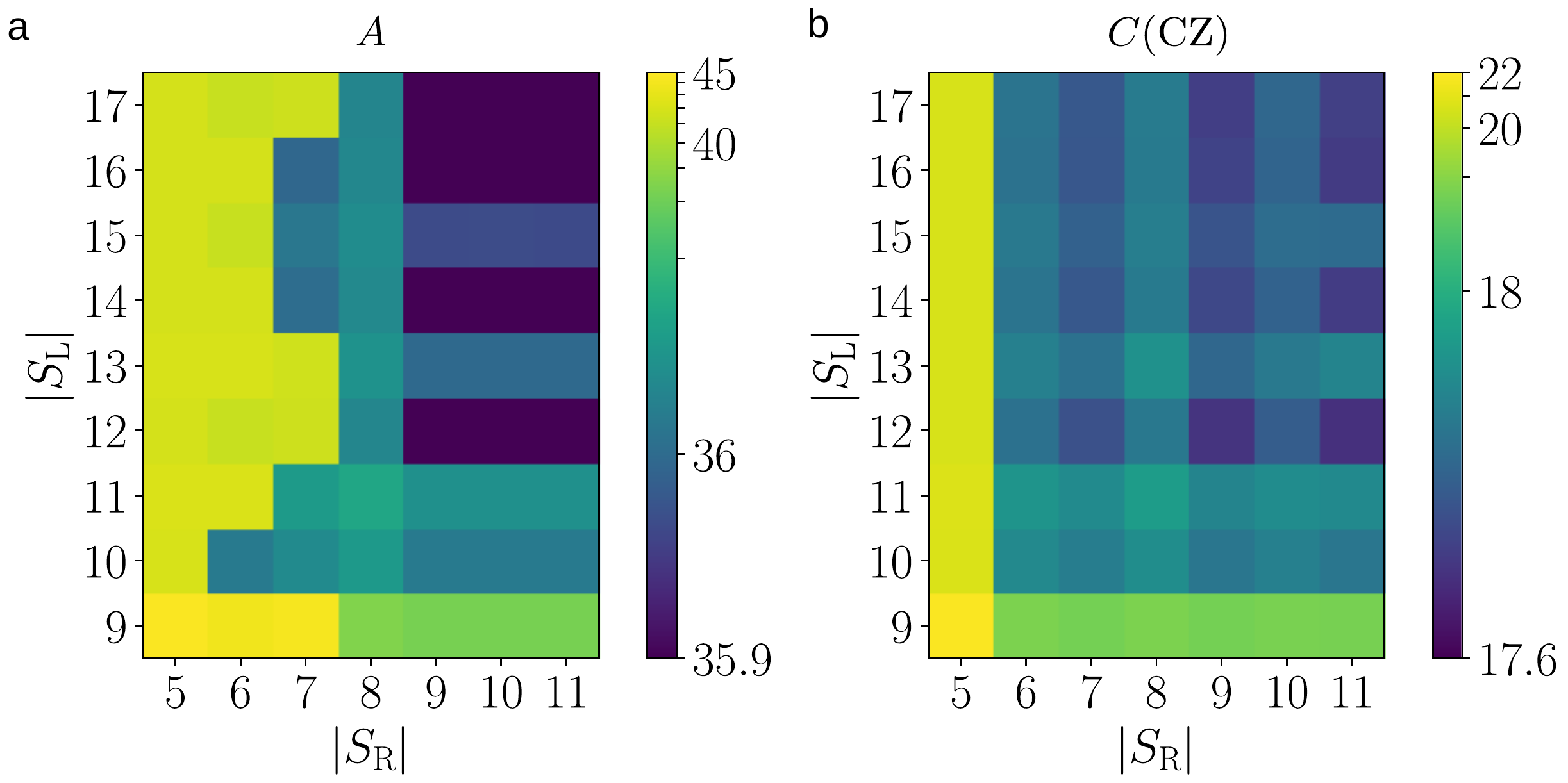}
    \caption{The optimized sample complexities with respect to different sample sizes. Notice that the complexity does not necessarily monotonically decrease. (a) The optimized upper bound $A$. There is a local minimum at $|S_L|=12$ and $|S_R|=9$ with $A=35.9$. (b) The optimized sample complexity for channel $\mathrm{CZ}$. There are local minima at $|S_L|=12$ and $|S_R|=9,11$ with $C(\mathrm{CZ}) = 17.6$. Using the set with $|S_L|=12$ and $|S_R|=9$ optimized in (a), we obtain $C(\mathrm{CZ})=17.9$.}
    \label{fig:qptac}
\end{figure}

The sets $S_L$ and $S_R$ are parametrized by the angles $\phi_i$ and $\alpha_i$, where $\theta_i = \pi/2$ are fixed because they are sufficient to rotate state $\ket{0}$ to arbitrary qubit state, and we use the gradient descent algorithm to minimize loss function channel-independent $A$ or channel-dependent $C$ with given set sizes. The minimization algorithm is run over various combinations of set sizes (Figure \ref{fig:qptac}), and we noticed a local minimum at $|S_L|=12, |S_R|=9$. The other observation is that, the sets obtained by optimizing $A$ usually performs as well for a specified channel. For example, the sample complexity $C(\mathrm{CZ})$ using sets $S_L$ and $S_R$ of sizes 12 and 9 obtained from minimizing $A$ is only $1.5\%$ greater than that using the sets optimized over $\mathrm{CZ}$ channel. 

Table \ref{tab:paramlist} gives the optimized sets $|S_L|$ and $|S_R|$.

If there are imperfections in the rotations, the algorithm will lead to a different convergence value from $\Lambda_{\mathcal{E}}$. Qualitatively, errors in rotations can be viewed as an additional term multiplied on $\mathcal{E}$. In the numerical simulation, we perturb the parameters $\alpha$ and $\phi$ to simulate the imperfections. The perturbations are either fixed, in which case we call them biased imperfections, or drawn randomly from a normal distribution $\mathcal{N}(0,\sigma)$, in which case we call them unbiased imperfections. For biased imperfections, we first generate the parameters $\alpha$ and $\phi$ from a normal distribution $\mathcal{N}(0,\sigma)$ which remain fixed. Notice that the biased imperfections give rise to predicted error linear in $\sigma$ due to the first order perturbation, while the unbiased imperfections give rise to predicted error quadratic in $\sigma$ from the second order perturbation.

\begin{table}[tbp]
    \centering
    \begin{tabular}{|c|cc||c|cc|}
    \hline
        $U_L^i$ & $\phi$ & $\alpha$ & $U_R^j$ & $\phi$ & $\alpha$\\
    \hline
        1 & -0.97332525 & -1.6321929 &  1 & -0.6033298 & -0.8976\\
        2 & 1.2753034 & 1.2730621 &  2 &  0.4034208 & -1.0890453 \\
        3 & -3.0044696 & 1.7525117 &  3 & -0.5442823 & -1.6806507 \\
        4 & 2.136892 & -1.1476487 &  4 &  1.7164205 & -3.068044  \\
        5 & 1.2600493 &  -2.2312932 &  5 & 1.401631 &  -0.786379  \\
        6 & -2.060111 &  1.136925  &  6 & 1.3496765 &  4.1021175 \\
        7 & 0.35845152 & 0.28960443 &  7 & 1.6976808 & -1.4610567 \\
        8 & -2.3170085 & -2.2987812 &  8 & 0.58622277 & -1.7876421 \\
        9 & -0.6583882  & 2.2054317 &  9 & 2.9298584 &  2.360425  \\
        10 & 0.13712281 & 1.3890826 &&&\\
        11 & 2.8700511 &  0.7049671&&& \\
        12 & 2.4798062 &  2.717094 &&& \\
    \hline
    \end{tabular}

    \caption{The optimized rotational unitaries for quantum process tomography (Algorithm I). The angles $\phi$ and $\alpha$ describing the rotations are defined in Eq.~\ref{eq:rotations}, and $\theta = \pi/2$.}
    \label{tab:paramlist}
\end{table}

\begin{figure}[htbp]
    \centering
    \includegraphics[width=0.55\linewidth]{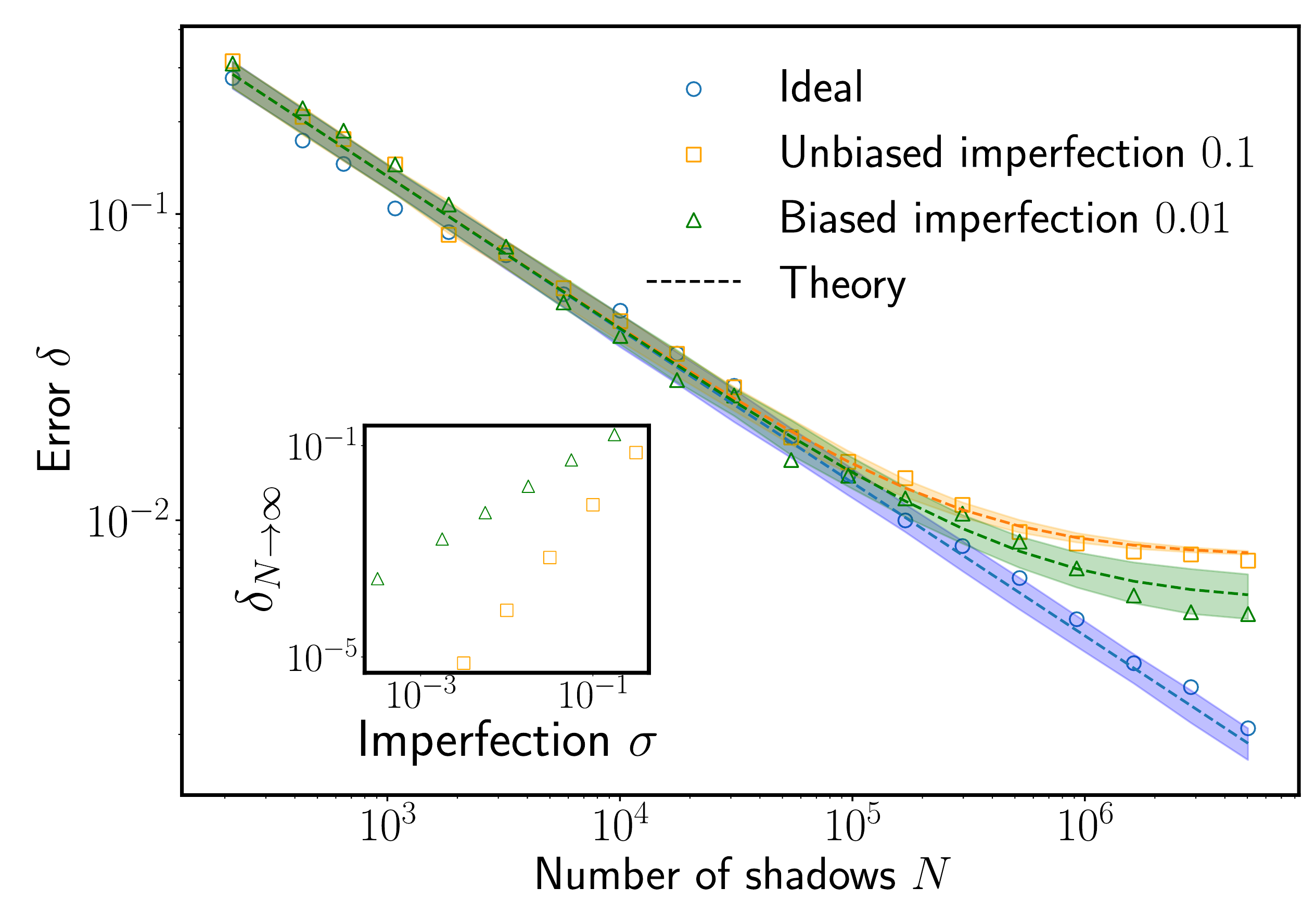}
    \caption{The convergence plot with phase errors only. The imperfection $\sigma$ is defined as in Eq.~\ref{eq:dephasing}.}
    \label{fig:a3}
\end{figure}

Figure~4(b) in the main text summarizes the results from simulating the effects of imperfections assuming both $\alpha$ and $\phi$ contain the same fluctuation. If there is only unbiased fluctuations in the rotation angle $\alpha$ of $U=e^{-i\alpha \mathbf{n}\cdot \mathbf{J}}$, then effectively the rotation can be viewed as a dephasing channel
\begin{equation}\label{eq:dephasing}
    \rho_f = (1-\sigma^2) U \rho U^\dagger + \sigma^2 \sigma_{\mathbf{n}} U \rho U^\dagger \sigma_{\mathbf{n}}
\end{equation}
where $\sigma_{\mathbf{n}}$ is the Pauli matrix along the rotation axis $\mathbf{n}$. It is clear that the error $\delta$ is proportional to $\sigma^2$. Figure~\ref{fig:a3} shows the convergence including only phase errors.

\end{document}